\documentclass[twocolumn]{aastex63}
\usepackage{graphicx}
\usepackage{natbib} 
\usepackage{url} 

\begin{document}

\def\spacingset#1{\renewcommand{\baselinestretch}%
{#1}\small\normalsize} \spacingset{1}





\title{Dark Matter Accounts for Perturbation in the GD-1 Stellar Stream}

\author{Carissma~McGee}
\affiliation{Howard University $|$ 2400 6th St NW, Washington, DC 20059}
\affiliation{Center for Astrophysics $|$ Harvard \& Smithsonian, 60 Garden Street, Cambridge, MA 02138, USA}

\author[0000-0002-7846-9787]{Ana~Bonaca}
\affiliation{Center for Astrophysics $|$ Harvard \& Smithsonian, 60 Garden Street, Cambridge, MA 02138, USA}

\begin{abstract}\noindent

The longest recognized stellar stream in the Milky Way Galaxy has an expanse of over more than half the north sky. There was a physical disturbance within the stream, 500 million years ago, which could have been the scar of a dark matter collision. Due to its proximity to the galactic center, the GD-1 stellar stream can act as an antenna for gravitational perturbations. In 2018, a significant gap in GD-1 was discovered due to perturbation. A stream gap occurs when a massive object collides with the stellar stream. Based on the chasm's location and width, we can guess when and where the impact occurred. Using globular cluster sky coordinates and simulated galactocentric distributions, we calculated how close each globular cluster came to the GD-1 stream. This would be the first time a globular cluster has come close enough to the GD-1 stream to impact another object. Clusters, on the other hand, rarely approach GD-1, indicating that it was struck by something more exotic, like a clump of dark matter, when discovered. A simulation and theoretical model were created to better understand the GD-1 stellar stream behavior. These details can be used to map the large-scale distribution of dark matter in our galaxy as well as the small-scale structure of dark matter in the host galaxies of the streams. Examination of stellar streams and detection of subhalos will not only confirm the presence of dark matter but also reveal information about its particle nature.

\end{abstract}

\noindent%

\spacingset{1.8} 
\section{Introduction}
\label{sec:intro}

Since the dawn of humanity, determining the number of stars visible in the night sky and forecasting their behavior has been a subject of interest. Timocharis of Alexandria and Hipparchus of Nicaea were among the first astronomers to begin counting and cataloguing naked-eye stars, creating the first magnitude-limited catalogues. Modern astronomers prefer catalogues with varying maximum distance limits, as any sample with a fixed magnitude is biased against intrinsically faint (and single) objects. We now know that they account for a substantial portion of the objects in our Galaxy, despite the fact that even the brightest of them are invisible to the naked eye.

Astronomers have been attempting to map our nearest neighbors since the first measurements of stellar parallaxes. Gaia, the second astrometric space mission, significantly improves the quality and quantity of sky parallax measurements for about 1.5 billion objects \cite{2016A&A...595A...1G}. It enables the completion of volume-restricted samples while increasing the distance limit. The Gaia Catalogue of Nearby Stars (GCNS), which is based on Gaia Early Data Release 3 (Gaia EDR3, Gaia Collaboration 2021a) \cite{2021arXiv210802531K}, extends the distance limit to 100 parsecs \citep{2016A&A...595A...1G}. Gaia determined all of these extremely precise positions and velocities, which we will use to map the stellar streams, identify perturbations, and diagnose their causes. Gaia determined all of these extremely precise positions and velocities, which we will use to map the stellar streams, identify perturbations, and diagnose their causes.

The Milky Way Galaxy's structure is fairly typical of a large spiral system \cite{2021PDU....3300838C}. This structure is composed of a nucleus, a central bulge, a disk, spiral arms, a spherical component, and a massive halo. Several of these components are virtually indistinguishable. At the heart of the Galaxy is an extraordinary object—a massive black hole encircled by a disk of hot gas known as an accretion disk. Due to the Milky Way's dense screen of intervening dust, optical wavelengths cannot be used to observe the central object or any of the material immediately surrounding it \cite{2021arXiv210608818B}. The nucleus is surrounded by an extended bulge of nearly spherical stars, the majority of which are Population II stars, albeit with a high concentration of heavy elements. Numerous globular clusters of similar stars are interspersed among the stars, and both the stars and the clusters orbit the nucleus in nearly radial orbits \cite{2021MNRAS.505.5978V}. Bulge stars can be seen optically where they rise above the obscuring dust of the galactic plane. The Milky Way halo contains a number of narrow, cool stellar streams that were formed as a result of tidal stripping of low-mass star systems such as global clusters \cite{2021gcf2.confE..63M}. These streams are critical for testing dark matter theories because they are used to map the Galaxy's overall mass structure and are highly sensitive to gravitational disturbance interactions with dark matter \cite{2021arXiv210802217W}. As a collection of tidally stripped dwarf galaxies and globular clusters, stellar streams provide excellent probes of the Milky Way's gravitational potential and can be used to precisely reconstruct the galaxy's past history.

Globular clusters are dense, symmetrical clusters of stars that primarily orbit within the extended star halos that surround the majority of spiral galaxies \cite{2021arXiv210705667B}. Globular clusters are believed to contain some of the galaxy's oldest stars and formed early in the galaxy's history.

Stellar streams are objects that enter the Milky Way and are unraveled by its tidal forces. As a result, they are extremely useful for tracing the hierarchical formation of the Milky Way galaxy's structure \cite{2021arXiv210608818B}. GD-1 is an unusually diverse stellar stream, prompting speculation about alternative processes that could account for various aspects of the data. Stellar streams are collections of stars that arc around their host galaxy in the form of elongated filaments \cite{2021nova.pres.7760K}. These filaments are thought to form when the tidal forces of the host galaxy disrupt a stream progenitor, such as a globular cluster or a satellite dwarf galaxy.

Although the nature of dark matter is unknown, it appears to account for roughly 82 percent of all matter in the universe. As a result, dark matter's gravitational interactions have shaped the evolution of the universe's structure \cite{2021DDA....5240106B}. Each galaxy is surrounded by an enigmatic halo of dark matter, which can be detected indirectly by observing its gravitational effects. The invisibly large and spherical halo dwarfs the brilliant galaxy at its center. Recent computer simulations have revealed a surprisingly clumpy halo, with relatively dense concentrations of dark matter contained within gravitationally bound'subhalos' \cite{2021arXiv210512131M}. The primary focus of this investigation will be on these "subhalos."

There is a great deal of uncertainty about the origins of the Milky Way's stellar streams \cite{2021gcf2.confE..63M}. Understanding the origins of these stellar streams enables us to more precisely trace their paths, the duration of their orbits, and any other gravitational interactions they may have encountered over time. We will develop a simulation and theoretical model to reconstruct a more accurate depiction of the GD-1 stellar stream behavior. These details are useful not only for understanding galaxy evolution, but also for mapping the big-picture distribution of dark matter in our galaxy and studying the small-scale structure of dark matter in the host galaxies of the streams \cite{2021arXiv210802217W}. Closer analysis of stellar streams along with detecting these subhalos will not only be a strong indicator for the existence of dark matter but also will give valuable information about its particle nature.

We are also interested to see if a dark-matter subhalo could have impacted the GD-1 stream. This is of course hard to determine right now, because we have not confirmed if they even exist in the time and space of our data set. However, we have an idea where they could be. Simulated evidence of old galaxy merging with the Milky Way closely fits the distribution of stars in the Milky Way halo \citep{naidu:2021}. This simulation also predicts where dark matter from this galaxy should be, so we can take these predictions and test if those subhalos could come close to GD-1 and other streams.



\section{Data}
\subsection{GD-1 Stream}

In the beginning, stellar streams are self-bound objects that fall into the Milky Way and are unravelled by the tidal forces of the Milky Way's potential \cite{2021arXiv210512131M}. As a result, they are extremely useful for tracing the hierarchical formation of the Milky Way galaxy's structure. GD-1 is a stellar stream with a surprising number of characteristics, which has sparked speculation about additional processes that could be used to explain various aspects of the data. 

By using uniform galactocentric coordinate structure throughout, I was able to make my own stream impact visualization in Figure 1. Although quantitative predictions for the kinematics of a spur formed by these processes have not been made, significant velocity kicks are expected.

\begin{figure}
\begin{center}
\includegraphics[width=3in]{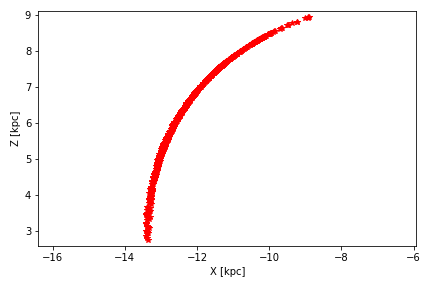}
\end{center}
\caption{Arc of the GD-1 Stellar Stream via galactocentric coordinates. \label{fig:first}
}
\end{figure}

We used the same Galactocentric coordinate system for the GD-1 stream, globular clusters, and simulated dark matter. 

This frame allowed for specifying the Sun-Galactic center distance, the height of the Sun above the Galactic midplane, and the solar motion relative to the Galactic center. However, as there is no modern standard definition of a Galactocentric reference frame. Once the specific stream impact was located, the next step was to analyze the distribution of Globular Clusters and check whether they passed through that area.

\subsection{Globular Cluster Distribution}

\begin{deluxetable}{c c c c c c}
\tablehead{ $X$ & $Y$ & $Z$ & $V_X$ & $V_Y$ & $V_Z$ \\ 
kpc & kpc & kpc & $\rm km\,s^{-1}$ & $\rm km\,s^{-1}$ & $\rm km\,s^{-1}$ } 
\decimals 
\setlength{\tabcolsep}{3pt} 
\startdata 
-6.3 & -2.6 & -3.2 & -75.0 & 165.1 & 46.0 \\ 
-8.2 & 0.0 & -8.9 & -7.0 & -52.0 & 51.3 \\ 
-5.0 & -5.1 & -6.2 & -88.2 & -100.3 & -69.4 \\ 
-22.2 & 4.7 & -26.2 & 227.7 & 54.2 & 8.2 \\ 
-8.1 & -10.0 & -12.8 & 77.3 & 50.9 & 69.1 \\ 
-14.9 & 8.0 & 3.7 & 65.3 & 197.7 & -19.4 \\ 
-24.9 & -80.1 & -92.2 & 120.0 & -71.2 & 105.6 \\ 
-61.5 & -41.8 & -59.3 & 92.0 & 16.9 & 133.6 \\ 
-34.6 & 4.4 & -4.2 & 109.9 & -13.5 & 3.8 \\ 
-12.4 & -8.9 & -6.9 & -83.4 & -72.8 & -81.1 \\ 
-15.8 & -8.3 & -6.3 & -44.1 & -21.1 & 5.0 \\ 
-12.4 & -9.5 & -2.9 & 94.9 & 19.5 & 77.7 \\ 
-82.7 & -0.5 & 35.4 & 6.1 & 36.5 & -57.5 \\ 
-14.0 & -38.7 & 4.8 & 114.9 & 211.2 & 187.1 \\ 
-6.1 & -9.2 & -1.9 & 57.2 & 142.5 & 28.9 \\ 
-5.2 & -7.1 & -2.6 & -224.9 & 106.5 & 105.0 \\ 
-42.3 & -59.7 & 61.8 & 15.2 & 160.3 & 67.3 \\ 
-7.5 & -4.8 & 0.8 & 258.1 & -205.4 & 150.6 \\ 
-39.3 & -12.9 & 103.4 & 36.3 & -34.5 & 56.1 \\ 
0.3 & -97.0 & 107.5 & 69.8 & 80.7 & 54.6 \\ 
-9.3 & -4.1 & 18.8 & -39.8 & -31.3 & 128.3 \\ 
-5.2 & -4.9 & -1.0 & -99.6 & 76.2 & 68.1 \\ 
2.5 & -17.8 & 4.3 & -120.4 & 217.6 & 33.3 \\ 
-4.0 & -7.2 & 6.1 & -167.9 & 282.2 & 16.7 \\ 
 
\enddata 
\caption{Distribution of globular clusters in the 6D phase space via Gaia Data Release 2} \label{table:dm} 
\end{deluxetable}

Globular clusters orbit in the halo of our galaxy, centered on the Galaxy’s center and expanding above and below the Galactic disk. Open star clusters tend to orbit within the disk. We analyzed the distribution of the globular clusters within 6D phase space utilizing both positions / velocity, and action / angle coordinates in combination with literature distance and line-of-view velocity measures. 

Globular star clusters contain hundreds of thousands of stars, and some -- like Omega Centauri –- contain millions of stars.  The population of the center 10 kpc clusters is 50-80 km/s rotating and 100-120 km/s isotropic, while the outer Galaxy shows that the cluster orbits are substantially anisotropical \cite{2021arXiv210411778T}. A proven combustion of clusters occurs in the outer portion of the Galaxy with high radial motion. Finally, the complete cluster system can explore a range of equilibrium distribution functions-based models coupled with information on the potential of the Milky Way. Models with circular velocity between 10 and 50 kpc stay in the 210-240 km/s range better reflect the dynamics of clusters.

The Gaia Release 2 data shown in Table 1 calculates the average proper motions of 150 Milky Way globular clusters (almost the whole known population), with a typical uncertainty of 0.05mas/yr constrained primarily by systematic errors \cite{2016A&A...595A...1G}. The dataset for Table 1 comes from a catalogue of mean proper motions of 150 Milky Way globular clusters, derived from the Gaia data release 2 astrometry, in a paper by Vasiliev. The columns are as follows: right ascension 2000, declination J2000, distance to the cluster, line-of-sight velocity, and proper motion right ascension.

\begin{figure}
\begin{center}
\includegraphics[width=3in]{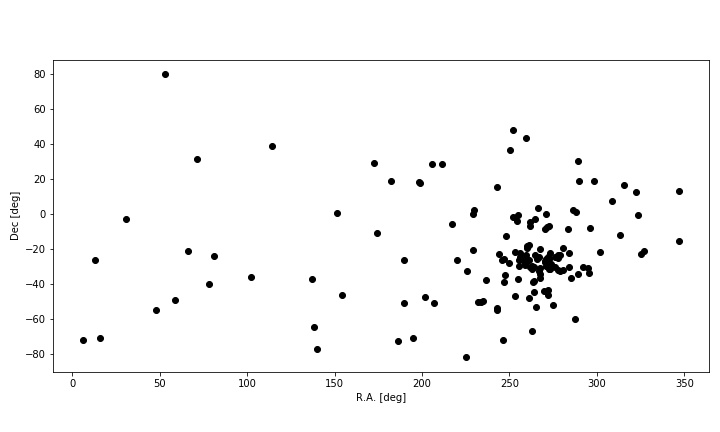}

\end{center}
\caption{The distribution of globular clusters in the 6D phase space, using both position/velocity and action/angle sky coordinates. \label{fig:second}
}
\end{figure}

\begin{figure}
\begin{center}
\includegraphics[width=3in]{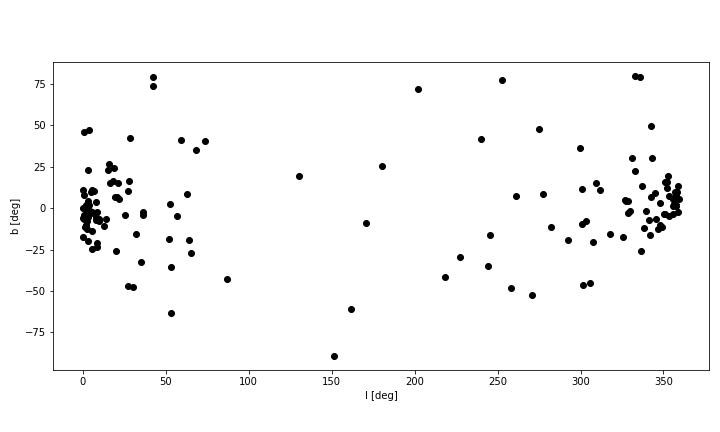}

\end{center}
\caption{The distribution of globular clusters in the 6D phase space, using both position/velocity and action/angle sky coordinates (cont.). \label{fig:third}
}
\end{figure}

In Figure 2 and Figure 3, the sky coordinates of globular clusters maintained the union of frame attributes for all built-in and user-defined coordinate frames in the astropy. Individual frame classes hold only the required attributes for that frame. The Sky Coordinates were analyzed on a three-tiered system of objects: representations, frames, and a high-level class. Representations classes are a particular way of storing a three-dimensional data point (or points), such as Cartesian coordinates or spherical polar coordinates. Frames are particular reference frames, which stored their data in different representations, but have well- defined transformations between each other. The Galaxy occupies a 3D region, hence this two-dimensional diagram only displays a piece of the plane of the Milky Way from top to bottom. Across various parts of the stream, clusters form along multiple dipoles.

\begin{figure}
\begin{center}
\includegraphics[width=3in]{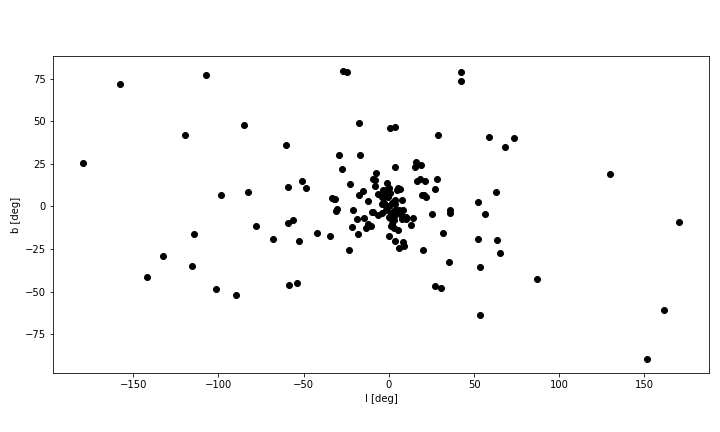}

\end{center}
\caption{Globular clusters sky coordinates . \label{fig:third}
}
\end{figure}

\begin{figure}
\begin{center}
\includegraphics[width=3in]{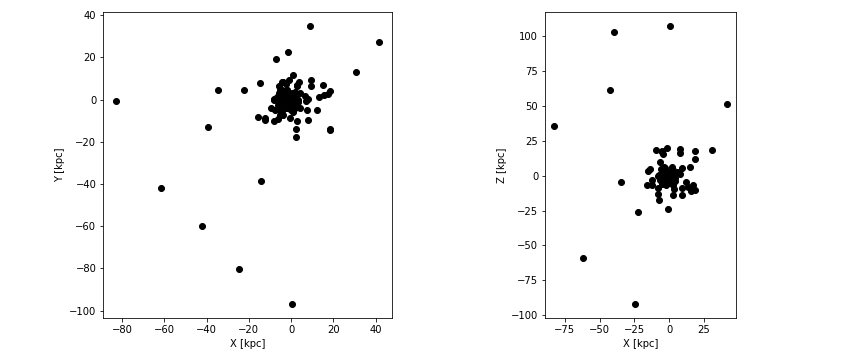}

\end{center}
\caption{X, Y, and Z coordinates of Globular Cluster distribution \label{fig:third}
}
\end{figure}

Figures 4 and 5, as well as variant observations of globular clusters in our range, can be explained by differences in the evolutionary paths of stars with similar compositions but different masses. The absolute magnitude at which the brighter main-sequence stars depart from the main sequence (the turnoff point, or "knee") indicates the cluster's age, assuming that the majority of the stars formed simultaneously. It is critical to note that globular clusters in the Milky Way Galaxy are nearly as old as the universe, averaging approximately 14 billion years and ranging between approximately 12 billion and 16 billion years, although these ages are still being revised.

\begin{figure}
\begin{center}
\includegraphics[width=3in]{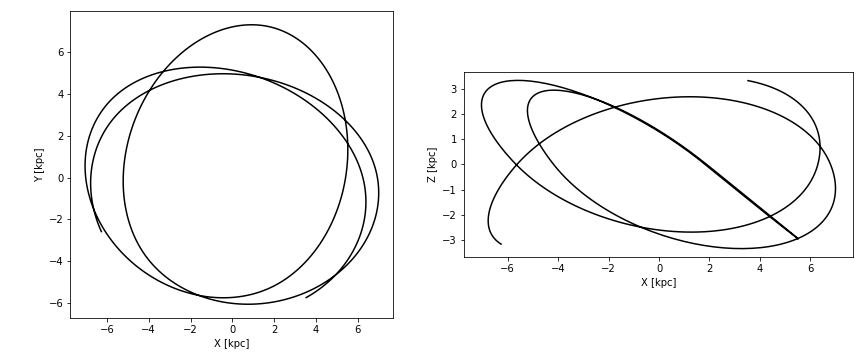}

\end{center}
\caption{The Hamiltonian orbit calculated by first finding the gravitational force and calculating how far it moves the object during a single time-step. \label{fig:third}
}
\end{figure}

\begin{figure}
\begin{center}
\includegraphics[width=3in]{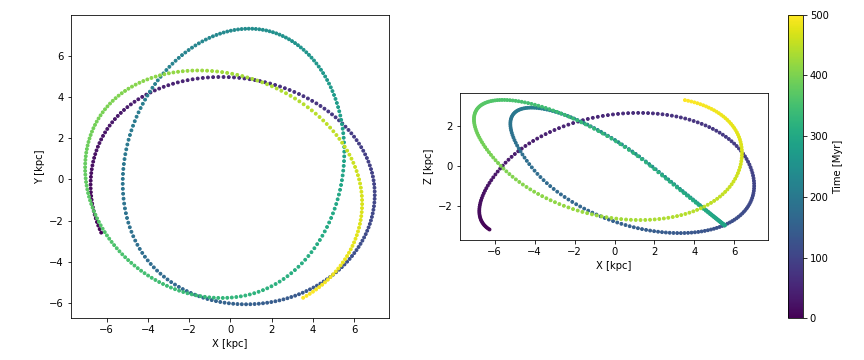}

\end{center}
\caption{A color magnitude plot of exactly where lot exactly where the globular clusters are at any given time. \label{fig:third}
}
\end{figure}

To perform the query and to retrieve the data, we used the Vasiliev
distribution to generate an orbital model from the globular cluster date, then we queried the Gaia science archive to download astrometric and kinematic data (parallax, proper motion, radial velocity) for a sample of stars near the Sun. Then we used data exclusively from data release 2 (DR2) from the Gaia mission module in the astroquery package, astroquery.gaia. By providing an SQL query to select our galactocentric coordinates and the sky position (ra, dec), parallax, proper motion components (pmra, pmdec), radial velocity, and magnitudes, we were able to visualize the orbit our of globular clusters in focus in Figure 6 and 7. 

\subsection{Dark Matter Distribution}


Due to their nearly universal ex situ origin, existing star streams are important indications of galaxy mergers and dynamical friction inside the Galactic halo. The stars revolving around the Milky Way's halo shed light on the distribution of dark matter in the galaxy, which had previously remained a mystery to researchers.

\begin{figure}
\begin{center}
\includegraphics[width=3in]{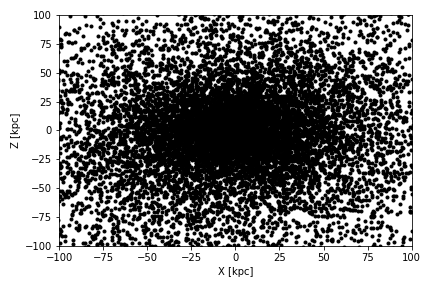}

\end{center}
\caption{Distribution Plot of Dark Matter Subhalo's via the Galactocentric Coordinate System. \label{fig:third}
}
\end{figure}

Figure 8 represents the distribution of dark matter sub halos in the exact same galactic coordinates as the orbit streams and globular clusters in the previous section. Their phase-space grouping allows for the creation of a precise global map of dark matter in the Milky Way, while their internal structure may provide insight into the small-scale structure of dark matter in their original home galaxies. 

\begin{deluxetable}{c c c c c c}
\tablehead{ $X$ & $Y$ & $Z$ & $V_X$ & $V_Y$ & $V_Z$ \\ 
kpc & kpc & kpc & $\rm km\,s^{-1}$ & $\rm km\,s^{-1}$ & $\rm km\,s^{-1}$ } 
\decimals 
\setlength{\tabcolsep}{3pt} 
\startdata 
-641.8 & 141.5 & -184.0 & -65.5 & -8.4 & 48.7 \\ 
47.7 & 22.3 & 84.2 & 69.7 & -61.4 & 3.5 \\ 
252.2 & -118.0 & 133.7 & 66.6 & 42.8 & -40.0 \\ 
-89.4 & 48.5 & -100.5 & -162.4 & -9.6 & -58.7 \\ 
-923.3 & 311.5 & -207.7 & -53.0 & 32.6 & -84.5 \\ 
-20.2 & 19.3 & -29.0 & -102.4 & 71.5 & -40.2 \\ 
235.4 & 42.2 & 3.1 & -31.9 & 78.4 & -69.0 \\ 
-218.5 & 742.1 & -152.5 & 7.2 & 118.1 & -11.7 \\ 
5.5 & 7.1 & -1.4 & -247.2 & -12.1 & -118.7 \\ 
-1101.3 & 11763.6 & 2334.8 & -104.2 & 46.7 & -73.5 \\ 
-172.3 & -43.5 & 67.3 & 4.1 & 96.6 & -51.8 \\ 
-18.4 & 29.6 & -68.3 & -137.8 & -104.9 & 81.8 \\ 
-44.4 & -6.2 & 9.3 & -88.7 & 62.0 & 28.1 \\ 
-1022.8 & -79.9 & 661.1 & -83.5 & 73.3 & -73.6 \\ 
-1546.3 & 590.3 & -2094.5 & -100.0 & 17.6 & -54.5 \\ 
-8.6 & -9.1 & -191.8 & -94.2 & 9.1 & -71.8 \\ 
198.8 & 429.2 & -467.7 & 51.3 & 51.2 & -50.9 \\ 
3.9 & 6.1 & 25.1 & 67.1 & 123.9 & 36.0 \\ 
-822.0 & -535.7 & -448.5 & -59.5 & 82.5 & -63.0 \\ 
-873.6 & 342.9 & -797.9 & -87.1 & -2.4 & -49.8 \\ 
-271.5 & 200.4 & -815.9 & -60.3 & 21.0 & -17.7 \\ 
3.0 & 26.2 & -42.7 & -82.7 & -161.3 & 97.9 \\ 
 
\enddata 
\caption{Present-day positions and velocities of simulated dark-matter particles in Galactocentric coordinates.} \label{table:dm} 
\end{deluxetable}

According to our investigation, the cosmologically motivated Dark Matter Halo profile was accompanied by reasonable baryonic component models (to account for the Milky Way's stellar bulge and disk) as well as reasonable dark matter component models (to represent the cosmos' dark matter halo). The flattening of the dark matter halo, as well as the circular velocity at the Solar radius, were used as proxy measurements for the Galaxy's total inner mass. Using the Galactic potential model, we varied these two parameters in order to estimate the GD-1 orbits. With this information, we determined that the GD-1 orbits must have a velocity of 244+/-4 kilometers per second at the solar radius and that the dark halo density must be in the range of 0.82+0.250.13 light years at the solar radius (a mildly oblate halo) \cite{2021ApJ...909L..26B}. According to one model, the Milky Way's mass is 2.5 billion solar masses and is located within 20 kpc of the Galactic Center. This provides the strongest support for that model. Recent studies have revealed that there is a greater degree of agreement among these values in the galactic halo, but there is still significant disagreement over the shape of the halo, which ranges from mildly oblate to mildly prolate in shape \cite{2021DDA....5240106B}.

\begin{figure}
\begin{center}
\includegraphics[width=3in]{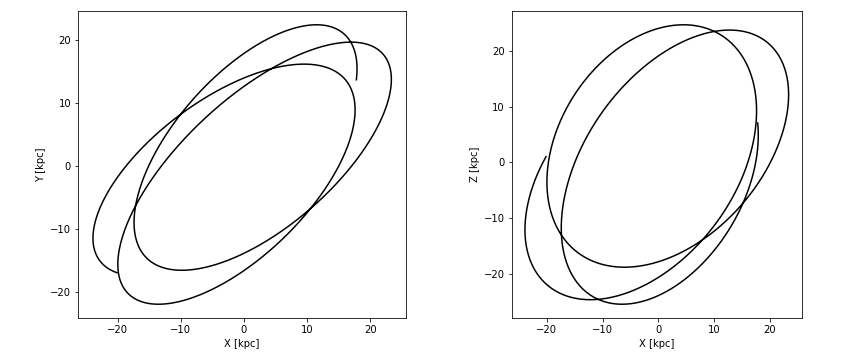}

\end{center}
\caption{The Horbit calculated by first finding the gravitational force and calculating how far it moves the Dark Matter Subhalo during a single time-step. \label{fig:third}
}
\end{figure}

The radial distribution of the subhalo population, shown within the orbits of in Fig. 9, have a slope equal to or shallower than the slope of the density profile for each resolution at all radii. At very small radii, subhaloes are highly deficient, possibly as a result, at least in part, of artificial disruption by the tidal forces from the central mass concentration. When a subhalo migrates inwards via dynamical friction to a radius where its central density is lower than the local density of the host halo, it will likely be disrupted. Numerical disruption enhanced by the strong tides near host centres could affect subhaloes. The flattened slope of the radial subhalo distribution of GRP1 subhaloes relative to the mass profile interior to roughly 0.2rvir for our highest resolution and interior to 0.3rvir for our lowest resolution implies that disruption and/or stripping of subhaloes is important in the halo central region. The increase in radius of the break in subhalo slope with decreasing resolution suggests that numerical disruption, if present, is worse for lower resolutions. However, we caution that the location of the break is not well defined because of Poisson uncertainties. Given the uncertainties, it is not possible to reliably separate spurious from real disruption that may be present in our simulations at small radii. Thus, we have no evidence that the central substructure number density has converged with resolution. Increasing the mass resolution by a factor of 8 results in roughly a factor of $\sim$2–2.5 more subhaloes at a given radius beyond roughly 0.3rvir, though there is substantial noise in this estimate. 

\section{Orbital Methods}


The Milky Way galaxy contains over 60 long and thin stellar streams \cite{2021ApJ...912...52G}. Due to their small size, these are expected to be tidally dissolved globular clusters, although low-mass dwarf galaxy progenitors are also permitted. Only a few streams are connected directly to a surviving globular cluster. The vast majority of streams have no apparent progenitor within them, and this section discusses how such streams formed.

In this section, we combine orbits for all of the Milky Way satellites (dwarf galaxies and globular clusters) in an attempt to link them to observed stellar streams. While we have shown that Milky Way streams and globular clusters are not associated with the plane of satellites as systems, it can still be interesting to associate individual VPOS members in order to try to establish a common origin (e.g. group infall). Associating streams with known objects, independent of the VPOS, can point to the stream's progenitor and trace the process of hierarchical formation.

In figure 6, we used the Galactocentric coordinates of the first globular cluster. This provided the data necessary to calculate the orbit and determine if the globular cluster was spending the same amount of time everywhere around the orbit. The Hamiltonian object we created before has a function associated with it that know to use the gravitational potential we specified already \cite{2021MNRAS.505.3033P}. By passing the initial position w0, we can view the size of the time-step (in our case 1 Myr), and the number of time-steps (in our case 500). The orbit is calculated by first finding the gravitational force and calculating how far it moves the object during a single time-step, then finding the gravitational force at the new position and repeating.

The full modeling of a satellite unraveling into a stellar stream is difficult because the results are dependent on the shape of the Galactic potential, the progenitor's internal and orbital kinematics, and the effects of dynamical friction and time-dependent substructures like the Galactic bar and the LMC \cite{2021MNRAS.505.3033P}. We integrate orbits in static potentials that do not include the effects of Galactic substructures or the LMC for simplicity and because we have limited information about each stream.

The initial positions for the globular cluster orbits are the same as in Section 1.2. Prior simulations of these initial positions, sampling over uncertainties in each distribution distance and proper motion were ran. We can get a sense of which pairings are possible given the measurement uncertainties by counting the fraction of orbits for each Milky Way satellite–stream pairing that are ‘associated'.


\section{Results}


\begin{figure}
\begin{center}
\includegraphics[width=3in]{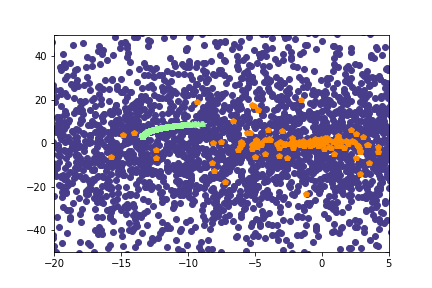}

\end{center}
\caption{Combined Diagram of the Stellar Stream, Globular Cluster, and Dark Matter Subhalo Distribution \label{fig:third}
}
\end{figure}

As a culmination of the stellar stream, globular cluster, and dark matter data from Gaia, we have demonstrated that the distributions of the Milky Way and stellar stream normals (Section 2.1) as well as the orbital poles of globular clusters (Section 2.2) are not significantly clustered in the direction of the VPOS normal. It may be tempting to use these results to completely reject the concept of a plane of satellites. We strongly advise against this, however. Taking the observed plane of satellite galaxies at face value, we must consider why other accreted matter – young halo globular clusters and stellar streams – does not exhibit the same behavior. Figure 10, reveals the expanse of the Milky Way examined in this project along with the data for each gala distribution. This plot is imperative to viewing distances from the stellar stream and determining if Dark Matter had an impact on the streams orbit, as well as ruling out globular clusters as a possibility. 

\begin{figure}
\begin{center}
\includegraphics[width=3in]{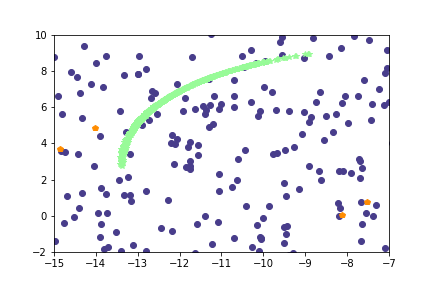}
\end{center}
\caption{Combined Diagram of the Stellar Stream, Globular Cluster, and Dark Matter Subhalo Distribution (Closer to the Stream). \label{fig:third}
}
\end{figure}

The orbits of previous integrations can be used to show that the stream encounter could not have been caused by any known globular cluster or dwarf galaxy, as they have been orbiting for the past 2 billion years. Also, it is assumed that the molecular cloud in the Milky Way disk was also mass, size, and impact parameter arguments that the stream encounter could not have been caused by. This hypothesized collision with a dark matter substructure, similar to those predicted to exist in galactic halos in LCDM cosmology, is the most plausible explanation for the gap-and-spur structure. While the GD-1 perturber density is higher than expected, the expected density of LCDM subhalos in this mass range and area of the Milky Way is lower. 

According to the results of orbit integrations performed back in time, the stream encounter could not have been caused by any known globular cluster or dwarf galaxy. In our opinion, the most likely explanation for our plot distribution is that the galaxy GD-1 was previously involved in an encounter with a clump of dark matter, such as those expected to exist in the galactic halo. If we can pinpoint the location of the perturber at this time, it will open the door to new research directions, including the search for additional observational evidence that the perturber is, in fact, dark matter. Other stars or gas clouds that are jostled around by the dark matter's gravity, or even gamma-rays associated with dark matter annihilations, which occur when two dark matter particles slam into and destroy each other, releasing a flash of energy, could be evidence of the existence of dark matter.

\begin{figure}
\begin{center}
\includegraphics[width=3in]{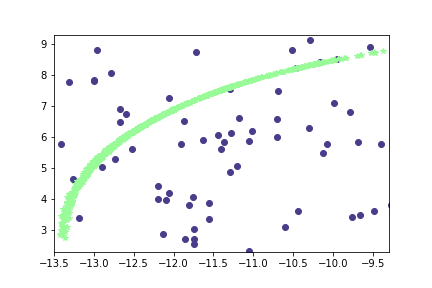}

\end{center}
\caption{Combined Diagram of the Stellar Stream, Globular Cluster, and Dark Matter Subhalo Distribution: Full Arc \label{fig:third}
}
\end{figure}

Extragalactic stellar streams may shed new light on low-mass galaxies \cite{2021ApJ...912...52G}. The progenitors of these streams are globular clusters that were previously a part of the globular cluster system of their host galaxies. Perhaps this recently dissolved population. On the other hand, assuming that streams formed recently implies that their host galaxies have larger pre-infall halo masses, many of which already have a sizable population of globular clusters. This is especially true for Sagittarius's accretion, as its impact on the Milky Way is highly dependent on its mass.

As illustrated in Figure 11 and 12, there are no globular clusters near the stream impact, which I obtained directly from Gaia Data. Our findings have broad implications for the use of stellar streams as dark matter tracers. To mention a few: (1) combining shorter streams into a single, longer structure increases their sensitivity to the gravitational potential's global properties; (2) high clustering of many stellar streams in phase space can be used to constrain the gravitational potential directly; (3) if the retrograde streams represent debris from a single merger event, they depict dynamical friction in action, whose magnitude would be much greater than what is visualized. 

We conclude by discussing the most urgently needed new stream observations. To extend this study to the entire population of streams in the Milky Way, high-quality proper motions of streams are required, many of which are remote and faint. Globular Clusters in the Milky Way. Our initial goal was to get familiar with globular clusters in the Milky Way. By visualizing globular clusters as point-mass objects, we were able to develop an improved understanding of possible perturbers.

\section{Discussion and Conclusions}
\label{sec:conc}

GD-1 is located between 8 and 10 kpc from the Sun, according to an analysis of the distances to various parts of the stream. There is only a shallow gradient across 45 degrees of the sky, according to the analysis \cite{2021MNRAS.505.2159B}. It is possible to see distinct density variations in the stream from a matched density map, including deviations from a single orbital track and preliminary evidence for stream fanning. Aside from that, we notice a distinct under-density at $\phi_1=-40$ deg in the center of the stream track, which is surrounded on either side by dense stream segments. 
It is clear that interactions with the Milky Way disk or other dark matter sub-halos have caused the GD-1 stream to be disrupted, as we have discovered.

Similarly, using our compilation of stream anchor points in Table 1, we discovered that stellar stream normals are not clustered around the normal, and that these findings are unaffected by geometric effects due to our observers' position.

We were successfully able to plot the sky positions of globular clusters (i.e., R.A. and Dec). We found asymmetric distribution, and when studying objects in the Milky Way, it was beneficial to convert the equatorial  sky  coordinates into  galactic  sky  coordi-nates, in whichl= 0, b= 0 points towards the Galactic center.  Using astropy coordinates, we converted the equatorial coordinates  of  the  Milky  Way  globular  clusters  into  galactic  coordi-nates and plot them. We discovered multiple possible instances of Milky Way globular clusters being associated with stellar streams using our orbit modeling procedure. According to our orbit modeling, it is quite possible that the globular clusters in the combined graphics in the previous section had a significant effect on the stream.

Finally, we reiterated our caution against using our findings to invalidate the concept of the plane of satellites. The satellite galaxies of the Milky Way have a highly clustered distribution of orbital poles within the spread figures of globular clusters, which contradicts expectations. 

Our findings, however, refute the notion that the Milky Way's satellite plane is a part of a composed of satellite galaxies, globular clusters, and stellar streams. They demonstrate that dark matter subhalos are the most likely explanation for the GD-1 stream impact. Our hope is that these findings will elicit interesting discussions about the stability of planar configurations and will serve as a springboard for future research on satellite planes in the context of hierarchical structure formation.

\section{Future Applications}

Known as dark matter, it is an intangible form of matter that accounts for the vast majority of the universe's mass and is responsible for the formation of galaxies. While astronomers are unable to directly observe dark matter, they can indirectly detect its presence by observing how the gravity of stars and galaxies is affected by the matter's gravitational pull \cite{2021CmPhy...4..123J}. It can be difficult, if not impossible, to detect the tiniest dark matter formations by looking for embedded stars because of the sparse star population in the region. Even in the absence of stars, the search for dark matter concentrations has proven to be problematic. Prior to Hubble's discovery, the Hubble research team used a technique that did not require them to look for the gravitational influence of stars as tracers of dark matter, which saved them time and money. In particular, the team concentrated on eight extremely bright and distant cosmic "streetlights," which they designated as quasars (regions around active black holes that emit enormous amounts of light). Astronomers measured the amount of distortion caused by the gravity of a massive foreground galaxy acting as a magnifying lens on the light emitted by oxygen and neon gas orbiting each quasar's black hole, according to the findings.

To associate a satellite with a stream, we can use gala to integrate the stream's orbit backwards in time. It is possible to examine whether the orbit of each object and stream pairing passes through the on-sky position of the stream's anchor points at approximately the correct heliocentric distance. An orbit is considered an association if it passes through both stream anchor points within the corresponding 1 angular and distance uncertainties. The next step after this duration of research is analyzing the Milky Way globular clusters and stellar streams for dynamical alignment with the Milky Way’s plane of satellites. 

Space telescope SPHEREX, which will be launched by NASA in 2024 and will be the first of its kind, may provide the lens needed to advance our understanding of Milky Way stream impact behavior. SPHERE is expected to launch no earlier than June 2024 and no later than April 2025 \cite{2021arXiv210712787C}. During its two-year mission, it will map the entire sky four times, resulting in a massive database of stars, galaxies, nebulas (space clouds of gas and dust), and a plethora of other celestial objects that will be available to researchers. With a total of 102 near-infrared colors observed, SPHERE will be the first NASA mission to create a full-sky near-infrared spectroscopy map, a first for the agency  \cite{2021arXiv210712787C}.

Large and medium-sized galaxies have been found to contain concentrations of dark matter, but until now, much smaller dark matter clumps have not been discovered in their immediate vicinity. Without any observational evidence for such small-scale clumps, some researchers have developed alternative theories, such as warm dark matter, to account for their observations of them \cite{2021PDU....3300838C}. According to this hypothesis, dark matter particles are extremely fast, zipping along at a rate that prevents them from merging and forming smaller concentrations of dark matter. It has been discovered that dark matter is colder than previously thought at smaller scales. There have been other observational tests of dark matter theories carried out in the past, but ours provides the strongest evidence to date for the presence of small clumps of cold dark matter in the universe. By combining the most recent theoretical predictions, statistical tools, and new observations, we can produce a more robust and up-to-date result than was previously possible through the theoretical simulation of this research.

\section{Acknowledgments}
I would like to acknowledge the support of Dr. Ana Bonaca who provided continued guidance, support, and a motivating spirit to explore every facet of the perturbation to analyze probable causes. I would like to acknowledge Rohan P. Naidu for providing data to help me better understand and curate my combined models.

The SAO REU program is funded in part by the National Science Foundation REU and Department of Defense ASSURE programs under NSF Grant no. AST1852268 and 2050813, and by the Smithsonian Institution.

\bibliography{references}
\bibliographystyle{aasjournal}

\end{document}